\documentclass[useAMS,usenatbib]{mn2e}

\def\deg{\tt $^{\rm o}$}
\def\arcmin{\tt '}

\title[Asteroseismology of $\nu$ Eridani -- IV]{Asteroseismology of the $\bbeta$ 
Cephei star $\bnu$ Eridani -- IV. The 2003-4 multisite photometric campaign and 
the combined 2002-4 data}

\author[M. Jerzykiewicz et al.]{ M.~Jerzykiewicz$^{1}$\thanks{E-mail:
mjerz@astro.uni.wroc.pl}, G.~Handler$^{2}$, 
R.~R.~Shobbrook$^{3}$\thanks{Visiting Fellow}, A.~Pigulski$^{1}$, 
R.~Medupe$^{4,5}$,\newauthor T.~Mokgwetsi$^{4}$, P.~Tlhagwane$^{4}$ 
and  E.~Rodr\'{\i}guez$^{6}$\\
$^{1}$Wroc{\l}aw University Observatory, ul. Kopernika 11, 51-622 Wroc{\l}aw, 
Poland\\
$^{2}$Institut f{\"u}r Astronomie, Universit{\"a}t Wien, 
T{\"u}rkenschanzstrasse 17, A-1180 Wien, Austria\\
$^{3}$Research School of Astronomy and Astrophysics, Australian National 
University, Canberra, ACT, Australia\\
$^{4}$Theoretical Astrophysics Programme, North West University, Private 
Bag X2046, Mmabatho 2735, South Africa\\
$^{5}$South African Astronomical Observatory, PO Box 9, Observatory 7935, 
South Africa\\
$^{6}$Instituto de Astrofisica de Andalucia, C.S.I.C., Apdo. 3004, 
18080 Granada, Spain}

\input epsf
\usepackage{graphics}
\begin{document}

\date{Accepted 2005 April 6. Received 2005 April 6; in original form 
2005 February 2}

\pagerange{\pageref{firstpage}--\pageref{lastpage}} \pubyear{2004}

\maketitle

\label{firstpage}

\begin{abstract}
The second multisite photometric campaign devoted to $\nu$ Eri 
is reported. The campaign, carried out from 11 Sept.\ 2003 to 16 Feb. 
2004, was very nearly a replica of the first, 2002-3 one: the five 
telescopes and photometers we used were the same as those in the first 
campaign, the comparison stars and observing procedure were identical, 
and the numbers and time base-lines of the data were comparable.  

For $\nu$ Eri, analysis of the new data adds four independent 
frequencies to the nine derived previously from the 2002-3 data, three 
in the range from 7.20 to 7.93  d$^{-1}$, and a low one, equal to 0.614 
d$^{-1}$. Combining the new and the old data results in two further 
independent frequencies, equal to 6.7322 and 6.2236 d$^{-1}$. 
Altogether, the oscillation spectrum is shown to consist of 12 high 
frequencies and two low ones. The latter have $u$ amplitudes about twice 
as large as the $v$ and $y$ amplitudes, a signature of high 
radial-order $g$ modes. Thus, the suggestion that $\nu$ Eri is both a 
$\beta$ Cephei and an SPB star, put forward on the basis of the first 
campaign's data, is confirmed. 

Nine of the 12 high frequencies form three triplets, of which two are 
new. The triplets represent rotationally split $\ell=1$ modes, although 
in case of the smallest-amplitude one this may be questioned. Mean 
separations and asymmetries of the triplets are derived with accuracy 
sufficient for meaningful comparison with models. 

The first comparison star, $\mu$ Eri, is shown to be an SPB variable 
with an oscillation spectrum consisting of six frequencies, three of which 
are equidistant in period. The star is also found to be an eclipsing 
variable. The eclipse is a transit, probably total, the secondary is 
fainter than the primary by several magnitudes, and the system is widely 
detached. 

The second comparison star, $\xi$ Eri, is confirmed to be a $\delta$ 
Scuti variable. To the frequency of 10.8742 d$^{-1}$ seen already in 
the first campaign's data, another one, equal to 17.2524 
d$^{-1}$, is added. 
\end{abstract}


\section{Introduction}

The first multisite photometric and spectrographic campaign devoted to 
the $\beta$ Cephei star $\nu$ Eridani was carried out between October 
2002 and February 2003. More than 600 h of differential $uvyV$ 
photometry on 148 nights and more than 2000 high-resolution spectra were 
obtained. A frequency analysis of the photometric data was reported by 
\citet[][hereafter Paper I]{hsj}, while the spectrographic time-series 
and their analysis were presented by \citet[][hereafter Paper II]{adh}. 
An extended frequency analysis and mode identification was provided by 
\citet[][hereafter Paper III]{dtb}.

Seismic modelling of the oscillation spectrum of $\nu$ Eri has been 
undertaken by \citet*{phd} and \citet{asta}.

In Paper I, the light-variation of $\nu$ Eri was shown to consist of 23 
sinusoidal terms. These included 8 independent ones with frequencies, 
$f_{i},\ i=1..8$, spanning the range from 5.6 to 7.9 
d$^{-1}$, 14 high-frequency combination terms, and a term with the low 
frequency $f_{\rm A}=0.432$ d$^{-1}$. 

The four highest-amplitude terms, discovered long ago by \citet{S76}, 
consist of a singlet with frequency $f_1=5.763$ d$^{-1}$, and a triplet 
very nearly equidistant in frequency ($f_3=5.620,\ f_4=5.637$ and 
$f_2=5.654$ d$^{-1}$). Before the campaign it was known that the singlet 
was a radial mode \citep*{cdp}, and surmised that it was the fundamental 
\citep{dj}, but the true nature of the triplet was unclear. From 
multicolour photometry, \citet*{hws} derived $\ell=1$ for $f_4$, and 
suggested that the triplet was a rotationally split dipole mode. This 
was challenged by \citet*{awp} who---from an analysis of line-profile 
variations---identified the $f_2$ term with an axisymmetric mode. 
However, both results are questionable because in neither case the 
triplet had been resolved. The matter was settled in Paper III: the 
wavelength dependence of the $uvy$ amplitudes of the triplet terms 
implies $\ell=1$ for all of them.  \citet{phd} showed then that the 
triplet was a $g_1$ mode. 

A frequency triplet $f_-<f_0<f_+$ can be characterized by its mean 
separation 
\begin{equation} S=0.5(f_+-f_-), 
\end{equation} 
and asymmetry 
\begin{equation} A=f_-+f_+-2f_0. 
\end{equation} 
In case of a rotationally split triplet, $S$ is determined by the 
angular rotation rate of the star, $\Omega$, while $A$ is sensitive to 
effects of higher order in $\Omega$, as well as to effects of a magnetic 
field. If these effects are negligible, $A=0$. 

For the $i=3,4,2$ triplet, both parameters were derived before the 
campaign from archival data by \citet{dj}. The value of asymmetry they 
obtained, $A=(-7.1\pm0.3)\times 10^{-4}$ d$^{-1}$, was unexpectedly 
large for the small rotation rate implied by the triplet's $S=0.0168$ 
d$^{-1}$. \citet{dj} showed that this problem may be solved by 
postulating the existence of a 5 to 10 kG magnetic field in the outer 
envelope of the star. From the campaign data (see Paper III), one gets 
$A=(-4.9\pm1.1)\times 10^{-4}$ d$^{-1}$, a value which is not in serious 
conflict with the observed $S$. Unfortunately, the large standard 
deviation of this result makes it useless. A longer time base-line than 
that of the 2002-3 campaign would be needed to obtain a more reliable 
value of $A$ and thus decide whether invoking magnetic field were 
necessary. This was one motivation for undertaking the sequel campaign. 

In Paper III, the spherical harmonic degree of the $i=5$, $6$ and $7$ 
terms was found to be $\ell=1$, but an attempt to derive $\ell$ for the 
low-frequency $i={\rm A}$ term (referred to as the $\nu_{10}$ term in 
that paper) was unsuccessful because of its small $uvy$ amplitudes and 
the poor resolution of diagnostic diagrams at low frequencies. 
\citet{phd} showed that the $i=5$ term is a $p_2$ mode, while the $i=6$ 
one is a $p_1$ mode. Moreover, they suggested that the low-frequency 
term is an $\ell=1$, $m=-1$, $g_{16}$ mode. 

According to \citet{phd}, only the $i=1$, $2$, $3$ and $4$ modes (i.e., 
the radial fundamental mode and the $\ell=1,\ g_1$ triplet) are 
unstable in standard models. Modes with $f>6$ d$^{-1}$ (i.e., the 
$i=5$, $6$ and $7$ ones) and the low-frequency mode are stable. 
\citet{phd} demonstrate, however, that a fourfold overabundance of Fe in 
the driving zone would account for excitation of the high-frequency modes, 
and would make the low-frequency mode marginally unstable. 

It has been noted in Paper I that if the low-frequency term were indeed 
a high-order $g$ mode, $\nu$ Eri would be both a $\beta$ Cephei variable 
and a slowly pulsating (SPB) star. Unfortunately, $f_{\rm A}$ differs 
from the sixth order combination frequency $3f_1-3f_3$ by less than 
0.003 d$^{-1}$. This number is much larger than the formal error of 
$f_{\rm A}$, but is smaller than half the frequency resolution of the 
campaign data. Thus, the possibility that $f_{\rm A}$ is the combination 
frequency---although rather unlikely---cannot be rejected. Again, a 
longer time base-line would help to settle the issue. 

In addition to extending the time base-line, the sequel campaign was 
expected to double the number of data points and thus lower the 
detection threshold so that modes having amplitudes too low to be seen 
in the 2002-3 frequency spectra could be discovered from the combined 
data. 

Both comparison stars used in the 2002-3 photometric observations turned 
out to be variable. For the first, $\mu$ Eri (HD\,30211, B5\,IV, 
$V=4.00$), the analysis carried out in Paper I revealed a dominant 
frequency $f'_1=0.6164$ d$^{-1}$. (From now on we shall use a prime to 
denote frequencies of $\mu$ Eri.) Prewhitening with this frequency 
resulted in an amplitude spectrum with a very strong $1/f$ component, 
indicating a complex variation. Since the star is a spectroscopic binary 
with an orbital period $P_{\rm orb}=7.35890$ d \citep{h}, tests have 
been made to detect a non-sinusoidal signal with this period. 
Unfortunately, they were inconclusive. 

Taking into account the star's position in the HR diagram, the frequency 
$f'_1$ of the dominant variation, and the fact that the $u$ amplitude 
was about a factor of two greater than the $v$ and $y$ amplitudes, it 
was concluded in Paper I that $\mu$ Eri is probably an SPB star. 
However, it was also noted that instead of pulsation, a rotational 
modulation could be the cause of the dominant variation. 

In Paper I, the frequency analysis of de-trended differential magnitudes 
of $\mu$ Eri revealed a small-amplitude variation with a frequency 
$f_{\rm x}=10.873$ d$^{-1}$. It was suggested that the second comparison 
star, $\xi$ Eri (HD\,27861, A2\,V, $V=5.17$), may be responsible. 

The present paper reports the sequel photometric campaign, carried out 
from 11 Sept.\ 2003 to 16 Feb.\ 2004, and an analysis of the data of 
both campaigns. The 2003-4 observations and reductions are described in 
Sect.\ 2. Sect.\ 3 contains an account of the frequency analysis of the 
new data and a comparison of the results with those of Paper I. Low 
frequencies in the variation of $\mu$ and $\nu$ Eri from the $uvy$ data 
of the first campaign are re-examined in Sect.\ 4. Sect.\ 5 is 
devoted to frequency analysis of the combined, 2002-4 data. Finally, 
Sect.\ 6 provides a summary with an emphasis on clues for 
asteroseismology of the three stars, $\nu$, $\mu$ and $\xi$ Eri.

\section{Observations and reductions}

The observations were carried out with five telescopes on four 
continents. An observing log is presented in Table 1. Comparison of this 
table with Table 1 of Paper I shows that the new data are almost as 
extensive as the old ones. However, the 2003-4 $y$ data are less 
numerous than the $v$ and $u$ ones (see below). The time base-lines of 
the 2002-3 and 2003-4 sets are very nearly the same; they amount to 
157.9 and 158.5 d for the old and new data, respectively.

The five telescopes and photometers were the same as those in the 2002-3 
campaign. Thus, single-channel photoelectric photometers were used at 
all sites but Sierra Nevada Observatory (OSN), where a simultaneous 
$uvby$ photometer was used. At OSN, the observations were obtained with 
all four Str\"omgren filters, at Fairborn, Lowell and Siding Spring, 
with $u$, $v$ and $y$. At SAAO, the $y$ filter used in the 2002-3 
campaign has deteriorated to such a degree that it had to be discarded. 
Since no replacement was available, the data were taken with two 
filters, Str\"omgren $u$ and $v$, except that on his first two nights AP 
used Johnson filters B and V.

\begin{table*}
 \centering
 \begin{minipage}{150mm}
  \caption{Log of the photometric measurements of $\nu$ Eri in 2003-4. Observatories 
are listed in the order of their geographic longitude.}
  \begin{tabular}{@{}lrrlrrl@{}}
  \hline
  \hline
  Observatory & Longitude & Latitude & Telescope & \multicolumn{2}{c}{Amount of data} 
& Observer(s)\\
   &  &  &  & Nights & h\ \ \ \ &  \\
  \hline
  \hline
 Sierra Nevada Observatory & $-$3\deg $23$\arcmin & +37\deg $04$\arcmin & 0.9-m & 4\ \ \ 
&  10.64 & ER  \\
 Fairborn Observatory & $-$110\deg $42$\arcmin & +31\deg $23$\arcmin & 0.75-m APT & 
 51\ \ \  & 211.09 & $--$ \\
 Lowell Observatory & $-$111\deg $40$\arcmin & +35\deg $12$\arcmin & 0.5-m & 25\ \ \  
& 87.10 & MJ \\
 Siding Spring Observatory & +149\deg $04$\arcmin & $-$31\deg $16$\arcmin & 0.6-m 
& 26\ \ \   & 79.77 & RRS \\
 South African Astronomical Observatory & +20\deg $49$\arcmin & $-$32\deg $22$\arcmin 
& 0.5-m & 17\ \ \  & 92.13 & AP\\
 South African Astronomical Observatory & +20\deg $49$\arcmin & $-$32\deg $22$\arcmin 
& 0.5-m & 19\ \ \  & 48.52 & RM, TM, PT\\
  \hline
Total & & & & 142\ \ \  & 529.25 & \\
\hline
  \hline
\end{tabular}
\end{minipage}
\end{table*}

The comparison stars and observing procedures were the same as in the 
first campaign. 

The reductions, carried out separately for each of the three wavelength 
bands, $u$, $v$, and $y$ or $V$, consisted in (1)~computing heliocentric 
JD numbers for the mean epochs of observations, (2) computing the air 
mass for each observation, (3) correcting instrumental magnitudes of 
$\nu$, $\mu$ and $\xi$ Eri for atmospheric extinction with first-order 
extinction coefficients derived from the instrumental magnitudes of 
$\xi$ Eri by means of Bouguer plots, (4) forming differential magnitudes 
`$\nu$ Eri $-$ $\xi$ Eri' and `$\mu$ Eri $-$ $\xi$ Eri,' and (5) 
setting the mean light-levels of the differential magnitudes from 
different telescope-filter combinations to the same values. In step 3, 
second-order extinction corrections were not applied because no 
colour-dependent extinction effects could be detected in the uncorrected 
differential magnitudes (but see the last paragraph of Sect.\ 3.1).  In 
step 4, the magnitudes of $\xi$ Eri were interpolated on the epoch of 
observation of $\nu$ or $\mu$ Eri. In step 5, the mean-light levels 
for each telescope-filter combination were derived using residuals from 
least-squares solutions with the four highest-amplitude terms (see the 
Introduction).

\section{Frequency analysis of the new data}

The analysis was carried out in essentially the same way as in Paper I. 
That is, frequencies were identified from periodograms, one at a time. 
Before each run, the data were prewhitened with all frequencies found in 
previous runs. After several runs, the frequencies were refined by means 
of a nonlinear least-squares fit using the values of independent 
frequencies read off the periodograms as starting values. The 
frequencies of the combination terms were computed from the independent 
frequencies. Thus, the unknowns in the normal equations were the 
corrections to the independent frequencies, to the mean magnitude, 
$\langle \Delta m \rangle$, and to the amplitudes, $A_{i}$, and phases, 
$\phi_{i}$, appearing in the following expression: 

\begin{equation}
\Delta m =  \langle \Delta m \rangle + \sum_{i=1}^N A_{i} 
\sin [2 \pi f_{i} (t - t_0) + \phi_{i}], 
\end{equation}
 where
 $\Delta m$ is the differential magnitude in $u$, $v$, or $yV$,
 $N$ is the number of all frequencies, $f_i$, the combination terms included, and 
 $t_0$    is an arbitrary initial epoch.

The difference with respect to the 2002-3 analysis consisted in using 
different programs: the PERIOD 98 package (see Paper I) was replaced 
with programs that have been in use by MJ since 1975 (see, e.g., 
\citealt{j78}). Thus, by ``periodogram'' we now mean a power spectrum, 
and not an amplitude spectrum as in Paper I; by ``power'' we mean 
$1-\sigma^2(f)/\sigma^2$, where $\sigma^2(f)$ is the variance of a 
least-squares fit of a sine-curve of frequency $f$ to the data, and 
$\sigma^2$ is the variance of the same data. However, for the purpose of 
estimating signal-to-noise ratios (see below) we also computed amplitude 
spectra. 

\subsection{The programme star}

The data used for analysis were the differential magnitudes `$\nu$ Eri 
$-$ $\xi$ Eri.' Power spectra were computed independently from the $u$ 
and $v$ data. No power spectra were computed from the less numerous $yV$ 
data, but nonlinear least-squares solutions were carried out for all 
three bands, with the starting values of the independent frequencies  
for $y$ (henceforth we shall use ``$y$'' instead of ``$yV$'') taken from 
$v$. The OSN $b$ and SAAO $B$ data were not used.

In the $u$ and $v$ power spectra, 14 independent and 15 combination 
frequencies could be clearly seen. They are identified in the first 
column of Table 2. The numbers in the table are from nonlinear 
least-squares solutions. The values of the independent frequencies and 
their standard deviations, computed as straight means from the separate 
solutions for $u$, $v$ and $y$, are given in column 2 above the 
horizontal line. The combination frequencies, listed below the line, were 
computed from the independent frequencies according to ID in the first 
column; their standard deviations were computed from the standard 
deviations of the independent frequencies assuming rms propagation of 
errors. The amplitudes, $A_u$, $A_v$ and $A_y$, with the standard 
deviations, given in columns 3, 4 and 5, respectively, are from the 
independent solutions for $u$, $v$ and $y$. The last column  
lists the $v$-amplitude signal-to-noise ratio, $S/N$, defined as in 
Paper I, except that in Paper I the mean noise level was estimated in 5 
d$^{-1}$ intervals, while now we adopted 0.1 d$^{-1}$ intervals for 
frequencies lower than 3 d$^{-1}$, and 1 d$^{-1}$ for higher 
frequencies. In all cases, the amplitude spectra of the data 
prewhitened with the 29 frequencies of Table 2 were used for estimating 
the mean noise level.  

\begin{table*}
 \centering
 \begin{minipage}{125mm}
  \caption{Frequencies and amplitudes in the differential magnitudes 
`$\nu$ Eri $-$ $\xi$ Eri' from the 2003-4 data. Independent frequencies 
are listed in the upper part of the table. The combination frequencies 
are listed below the  horizontal line. In both cases the frequencies are 
ordered according to decreasing $v$ amplitude, $A_v$. The last column 
contains the $v$-amplitude signal-to-noise ratio. Frequencies $f_{\rm x}$ 
and $f_{\rm y}$ are due to $\xi$ Eri.}
  \begin{tabular}
{rl@{\hspace{3pt}$\pm$\hspace{3pt}}lr@{\hspace{3pt}$\pm$\hspace{3pt}}
rr@{\hspace{3pt}$\pm$\hspace{3pt}}rr@{\hspace{3pt}$\pm$\hspace{3pt}}rr} \\
  \hline
  \hline
  ID &\multicolumn{2}{c}{Frequency [d$^{-1}$]} & \multicolumn{2}{c}{$A_u$ [mmag]} & 
\multicolumn{2}{c}{$A_v$ [mmag]} & \multicolumn{2}{c}{$A_y$ [mmag]}&$S/N$\\
 \hline
 \hline
$f_1 $&\ \,5.763256&0.000012&72.3&0.19&40.6&0.14&36.7&0.15&165.0\\
$f_2 $&\ \,5.653897&0.000020&38.6&0.19&27.2&0.14&25.4&0.15&110.6\\
$f_3 $&\ \,5.619979&0.000021&35.5&0.19&25.0&0.14&24.0&0.15&101.6\\
$f_4 $&\ \,5.637215&0.000025&31.0&0.18&21.8&0.13&20.4&0.16&88.6\\
$f_5 $& \ \,7.89859&0.00022&3.7&0.19&2.7&0.14&2.5&0.16&11.0\\
$f_7 $& \ \,6.26225&0.00025&3.1&0.19&2.2&0.14&2.2&0.16&8.8\\
$f_{\rm A} $& \ \,0.43257&0.00028&3.2&0.19&2.0&0.14&1.7&0.16&5.0\\
$f_6 $& \ \,6.24468&0.00034&2.3&0.18&1.6&0.13&1.6&0.15&6.4\\
$f_{\rm B} $&\ \,0.61411&0.00033&3.4&0.18&1.5&0.13&1.6&0.15&4.3\\
$f_{\rm x} $&10.8743&0.0006&1.2&0.18&1.4&0.13&0.9&0.15&5.7\\
$f_{10}$& \ \,7.9296&0.0005&1.6&0.19&1.2&0.14&1.2&0.16&5.3\\
$f_9 $& \ \,7.9132&0.0005&1.8&0.18&1.1&0.13&1.3&0.16&4.9\\
$f_8 $& \ \,7.2006&0.0005&1.4&0.18&1.0&0.13&1.1&0.15&4.4\\
$f_{\rm y} $&17.2534&0.0006&1.0&0.19&0.9&0.13&0.9&0.15&4.6\\
\hline
$ f_1+f_2    $&11.417153 &0.000023 & 12.3&0.19 & 8.8&0.14& 8.4&0.15&34.0\\
$ f_1+f_3    $&11.383235 &0.000024 & 10.8&0.19 & 7.6&0.14& 7.1&0.15&29.3\\
$ f_1+f_4    $&11.400471 &0.000028 &  9.9&0.19 & 7.0&0.14& 6.6&0.16&27.0\\
$ 2f_1      $&11.526512 &0.000017 &  4.5&0.18 & 3.1&0.13& 3.1&0.15&12.0\\
$ f_1+f_2+f_3 $&17.037132 &0.000031 &  3.3&0.18 & 2.2&0.13& 2.0&0.15&11.3\\
$ f_1-f_2    $& \ \,0.109359&0.000023 &  2.7&0.19 & 1.6&0.14& 1.6&0.16&4.4\\
$ 2f_1+f_4   $&17.163727 &0.000030 &  1.9&0.18 & 1.4&0.13& 1.3&0.15&7.2\\
$ 2f_1+f_2   $&17.180409 &0.000026 &  1.9&0.18 & 1.4&0.13& 1.3&0.15&7.2\\
$ f_2+f_3    $&11.273876 &0.000029 &  2.4&0.19 & 1.3&0.14& 1.1&0.15&5.0\\
$ 2f_1+f_3   $&17.146491 &0.000027 &  1.4&0.18 & 1.2&0.13& 0.8&0.15&6.2\\
$ 2f_2      $&11.307794 &0.000028 &  1.1&0.19 & 1.1&0.14& 0.6&0.16&4.2\\
$ f_1-f_4    $& \ \,0.126041&0.000028 &  1.9&0.18 & 1.1&0.13& 1.2&0.15&3.0\\
$ f_1+f_3+f_4 $&17.020450 &0.000035 &  1.3&0.18 & 1.1&0.13& 0.9&0.15&5.6\\
$ 2f_1+f_2+f_3$&22.800388 &0.000034 &  1.3&0.18 & 1.1&0.13& 0.8&0.15&6.1\\
$ f_1+f_2-f_3 $& \ \,5.797174&0.000031 &  1.5&0.19 & 0.9&0.14& 1.3&0.15&3.7\\
\hline
\hline
\end{tabular}
\end{minipage}
\end{table*}

In Paper I, a peak in the amplitude spectrum was considered to be 
significant if its signal-to-noise ratio exceeded 4 for an independent 
term or 3.5 for a combination term.  As can be seen from Table 2, this 
condition is met by all terms identified from the 2003-4 data except the 
differential combination term $f_1-f_4$ for which $S/N=3.0$. This term 
may be spurious. 

The standard deviations in Table 2 (and in the tables that follow), 
referred to as ``formal'' in the remainder this subsection and in the 
first three paragraphs of the next subsection, will be underestimated if 
the errors of differential magnitudes are correlated in time. For the 
case at hand, i.e., of fitting a sinusoid to time series data, the 
problem has been discussed by \citet{s-c91} and \citet{mod}. These 
authors consider the factor, $D$, by which the formal standard 
deviations of a frequency, amplitude and phase should be multiplied in 
order to get correct values. ($D$ is the same for frequencies, 
amplitudes and phases because the ratio of the formal standard 
deviations of any two of these quantities is determined by 
error-free numbers such as the epochs of observations and the starting 
values of the frequencies.) The factor depends on an estimate of 
the number of consecutive data points which are correlated. 
Unfortunately, this estimate is not easy to obtain, especially for time 
series such as ours, consisting of data from many nights and several 
observatories. We shall return to this point in the next subsection. 

\begin{figure} 
\includegraphics{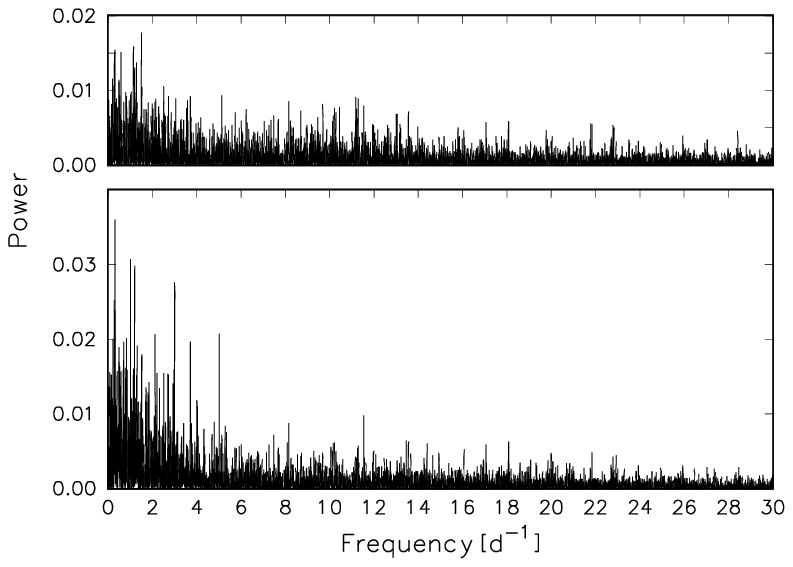} 
 \caption{Power spectra of the $u$ (lower panel) and $v$ (upper) 
differential magnitudes `$\nu$ Eri $-$ $\xi$ Eri' prewhitened with the 29 
frequencies of Table 2.}
 \label{Fig_1}
\end{figure}

Figure~\ref{Fig_1} shows the power spectra of the $u$ and $v$ 
differential magnitudes `$\nu$ Eri $-$ $\xi$ Eri' prewhitened with the 
29 frequencies of Table 2. In both panels of the figure, the highest 
peaks are seen at frequencies lower than about 5 d$^{-1}$. In the $u$ 
spectrum (lower panel), the highest peak occurs at 0.314 d$^{-1}$, while 
in the $v$ spectrum, the highest peak is the one at 1.498 d$^{-1}$. The 
$u$ amplitude at 0.314 d$^{-1}$ is equal to $2.2$ mmag, while the $v$ 
amplitude at 1.498 d$^{-1}$, to $1.1$ mmag. The $u$- and $v$-amplitude 
$S/N$ amount to 2.7 in both cases. We conclude that these frequencies 
are probably spurious. 

For frequencies higher than 5 d$^{-1}$, the highest peak in the $u$ 
spectrum (lower panel) occurs at 5.015 d$^{-1}$. For this peak the 
$u$-amplitude $S/N$ is equal to 5.0, so that the significance condition 
mentioned earlier in this section is satisfied. However, there is no 
power at 5.015 d$^{-1}$ in the $v$ spectrum (upper panel); the peak 
closest to 5.015 d$^{-1}$, at 5.128 d$^{-1}$, is probably noise because 
it has $S/N$ equal to 3.2. This, and the fact that 5.015 d$^{-1}$ is 
very nearly equal to 5 cycles per sidereal day (sd$^{-1}$), suggest that 
the peak is due to colour extinction in the $u$ band which we neglected 
(see Sect.\  2). Indeed, the differential colour-extinction correction 
contains a term equal to $k'' X \Delta C$, where $k''$ is the 
scond-order extinction coefficient, $X$ is the air-mass, and $\Delta C$ 
is the differential colour-index. Because of the second factor, $X$, the 
term causes a parabola-shaped variation symmetrical around the time of 
meridian passage. For a single observatory, this will produce a signal 
with frequency equal to $n$ sd$^{-1}$, where $n$ is a small whole number 
equal to the number of sinewaves that best fit the variations on 
successive nights. In our case this number is apparently equal to 5. For 
multi-site data, phase smearing may occur, tending to wash out the 
signal. However, our time-series is dominated by the data from Fairborn 
and Lowell, two observatories lying on nearly the same meridian and 
therefore introducing negligible phase shift. In addition, the greatest 
common divisor of the longitude differences between Fairborn and Lowell 
on one hand, and SAAO and Siding Spring on the other is close to $60^o$, 
making the 5 sd$^{-1}$ signals from these observatories to approximately 
agree in phase. We conclude that the 5.015 d$^{-1}$ peak in the $u$ 
spectrum in Fig.~\ref{Fig_1} is not due to an intrinsic variation of 
$\nu$ Eri. 

\subsection{Comparison of the 2003-4 and 2002-3 results for the programme 
star}

Of the 14 independent terms in Table 2, nine appear in Table 2 of Paper 
I. The differences between the 2003-4 and 2002-3 values of their 
frequencies and amplitudes are listed in Table 3. The standard 
deviations given in the table were computed from the standard deviations 
of the 2003-4 and 2002-3 least-squares solutions assuming rms 
propagation of errors. Thus, they are formal in the sense defined in 
the previous subsection. 

We shall now return to the factor $D$ by which the formal 
standard deviations are underestimated. Let us consider the nine 
independent frequencies common to 2002-3 and 2003-4. As can be seen from 
Table 3, the moduli of the frequency differences, $|\Delta f|$, are of 
the same order of magnitude as the formal standard deviations of $\Delta 
f$, $\sigma_{\Delta {\rm f}}$. In four cases $|\Delta f| < 
\sigma_{\Delta {\rm f}}$, in three cases $\sigma_{\Delta {\rm f}} < 
|\Delta f| < 2\sigma_{\Delta {\rm f}}$, and  in two cases the 
differences exceed $2\sigma_{\Delta {\rm f}}$. Taken at face value, 
these inequalities would indicate that four or five of the nine 
frequencies changed from 2002-3 to 2003-4 by 1$\sigma_{\Delta {\rm f}}$ 
or more, while $f_3$, $f_5$ and $f_8$ changed by 2$\sigma_{\Delta {\rm 
f}}$ or more. However, frequency changes in $\beta$ Cephei stars 
have time scales much longer than 1 year. For the four first frequencies 
of $\nu$ Eri, this is shown to be the case by \citet{hjpp}. Although the 
existence of long-term variations does not exclude the possibility of 
year-to-year ones, let us assume that these four frequencies were 
strictly constant from 2002 to 2004. If the assumption were false, the 
value of $D$ we derive in the next paragraph will be too large. 

For $f=const$, where $f$ is any of the four frequencies, the modulus 
of $\Delta f$ can be thought of as the range of $f$ in a two-element 
sample of $f$, the first element chosen from the parent population of 
$f$ in 2002-3, and the second element chosen from the same population in 
2003-4. For a normal distribution, an estimate of the standard deviation 
can be obtained by multiplying the range by a coefficient $k$ which is a 
function of the number of elements in the sample, $n$. For $n=2$, Table 
12 of \citet*{cdm} reads $k=0.886$. Multiplying $|\Delta f|$ by this 
value we get an estimate of the standard deviation of $f$. The latter 
number divided by the formal standard deviation of $f$ from Table 2 
yields the factor we are seeking.  The mean value of the factor for the 
four frequencies turns out to be $D=2.0$. If we applied the procedure to 
all nine frequencies, the result would be $D=1.8$. If we used the formal 
standard deviations of the 2002-3 solution, the results would be very 
nearly the same. Henceforth we shall adopt $D=2$. A standard deviation 
equal to the formal standard deviation times this factor we shall refer 
to as ``corrected,'' and from now on we shall drop the adjective 
``formal,'' so that by ``standard deviation'' we shall mean ``formal 
standard deviation.'' 

\begin{table*}
 \centering
 \begin{minipage}{105mm}
  \caption{A comparison of the frequencies and amplitudes of the nine 
independent terms common to Table 2 of Paper I and Table 2 of the present paper. 
The differences, $\Delta$, are in the sense `2003-4 $minus$ 2002-3.'}
  \begin{tabular}
{rl@{\hspace{3pt}$\pm$\hspace{3pt}}lr@{\hspace{3pt}$\pm$\hspace{3pt}}
lr@{\hspace{3pt}$\pm$\hspace{3pt}}lr@{\hspace{3pt}$\pm$\hspace{3pt}}l} \\
  \hline
  \hline
  ID &\multicolumn{2}{c}{$\Delta f$ [d$^{-1}$]} & \multicolumn{2}{c}
{$\Delta A_u$ [mmag]} & \multicolumn{2}{c}{$\Delta A_v$ [mmag]} 
& \multicolumn{2}{c}{$\Delta A_y$ [mmag]}\\
 \hline
 \hline
$f_1 $&$-$0.000014&0.000017&$-$1.2&0.27&$-$0.4&0.20&$-$0.2&0.20\\
$f_2 $&$-$0.000033&0.000028&0.7&0.27&0.8&0.20&0.3&0.20\\
$f_3 $&$-$0.000081&0.000030&0.9&0.27&1.0&0.20&1.3&0.20\\
$f_4 $&\hspace{7pt}0.000055&0.000035&$-$1.2&0.27&$-$0.6&0.20&$-$0.6&0.20\\
$f_5 $&\hspace{7pt}0.00079&0.00032&$-$0.6&0.27&$-$0.4&0.20&$-$0.5&0.20\\
$f_6 $&\hspace{7pt}0.00060&0.00048&$-$1.6&0.27&$-$0.9&0.20&$-$1.0&0.20\\
$f_7 $&\hspace{7pt}0.00020&0.00035&0.2&0.27&0.3&0.20&0.4&0.20\\
$f_8 $&\hspace{7pt}0.00066&0.00071&0.1&0.27&0.1&0.20&0.0&0.20\\
$f_{\rm A} $&\hspace{7pt}0.00039&0.00040&$-$2.3&0.27&$-$1.2&0.20&$-$1.5&0.20\\
\hline
\hline
\end{tabular}
\end{minipage}
\end{table*}

Multiplying the standard deviations of the 2003-4 amplitudes by two, we 
find that the ratios of the amplitudes to the corrected standard 
deviations become approximately equal to the signal-to-noise ratios 
defined in the previous subsection. This is a pleasant surprise, lending 
support to our value of $D$. (Using the numbers from Table 2, the reader 
can verify the approximate equality of $S/N$ and $A_v/(2\sigma_v)$, 
where $\sigma_v$ is the standard deviation of $A_v$. Note that while for 
low frequencies $A_v/(2\sigma_v)$ is about 30\% larger than $S/N$, the 
approximate equality improves for high frequencies. This seems to be 
reasonable in view of the $1/f$ decrease of the noise level in the  
periodograms.) Additional support for $D=2$ comes from the fact 
that a different line of reasoning applied to multisite data similar to 
ours has led \citet{has} to the same value.

Let us now return to Table 3. As can be seen from the table, $|\Delta 
A|$ are the largest for $f_{\rm A}$, $f_6$, and $f_3$ (in this order). 
Multiplying the standard deviations of $\Delta A$ by $D=2$ we find that 
(1) the amplitude of the low frequency term has decreased by 
4.3$\sigma_{\rm c}$ in $u$, 3.0$\sigma_{\rm c}$ in $v$, and 
3.8$\sigma_{\rm c}$ in $y$, where by $\sigma_{\rm c}$ we denote the 
corrected standard deviation of $\Delta A$, (2)~the amplitude of the 
$i=6$ term has decreased by 3.0, 2.2, and 2.5$\sigma_{\rm c}$ in $u$, 
$v$, and $y$, respectively, and (3)~the amplitude of the $i=3$ term has 
increased by 1.7, 2.5, and 3.2$\sigma_{\rm c}$ in $u$, $v$, and $y$, 
respectively. For the remaining six terms, $|\Delta A| \le 
2.2\sigma_{\rm c}$. We conclude that the decrease of the amplitude of 
the $i={\rm A}$ term is real, that of the $i=6$ term may be real, while 
the amplitude increase of the $i=3$ term is probably spurious. In the 
remaining cases, there is little or no evidence for amplitude variation. 

The five independent frequencies which appear in the present Table 2 but 
not in Table 2 of Paper I are (in the order of decreasing $v$ amplitude) 
$f_{\rm B}$, $f_{\rm x}$, $f_{10}$, $f_9$, $f_{\rm y}$. The low 
frequency $f_{\rm B}$ is very nearly equal to the frequency $f_{21}$, 
one of three low frequencies derived in Paper II from the radial 
velocities of $\nu$ Eri. However, in our 2003-4 power spectra we did not 
see the other two low frequencies of Paper II. Frequency $f_9$ was 
listed in Table 3 of Paper I as that of one of several ``possible 
further signals.'' Frequencies $f_{10}$ and $f_{\rm y}$ are new.

Frequency $f_{\rm x}$ was found in Paper I and tentatively ascribed to 
$\xi$ Eri (see the Introduction). We shall demonstrate in Sect.\ 3.3 
that this frequency and $f_{\rm y}$, the smallest-amplitude independent 
frequency in Table 2, are both due to $\xi$ Eri. 

In addition to nine independent terms, Table 2 of Paper I lists 14 combination 
terms. In the 2003-4 power spectra we did not see three of them, viz., 
$f_1+f_5$, $f_1+f_2+f_4$ and $f_1+2f_2$. On the other hand, we detected 
differential combination-terms $f_1-f_2$, $f_1+f_2-f_3$ and $f_1-f_4$ which 
were not found in Paper I.

\subsection{The comparison stars}

The data used for analysis were the differential magnitudes `$\mu$ Eri 
$-$ $\xi$ Eri.' Power spectra were computed independently from the $u$ 
and $v$ data. No power spectra were computed from the less numerous $y$ 
data.

\begin{figure}
\includegraphics{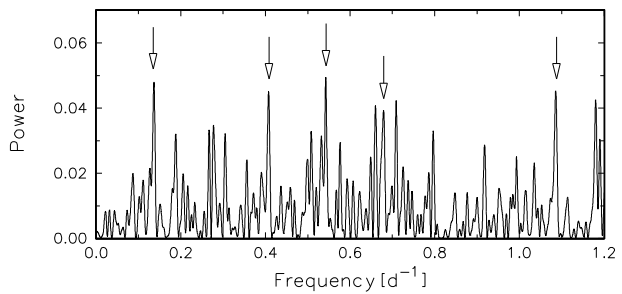}
 \caption{Low-frequency part of the power spectrum of the $v$ 
differential magnitudes `$\mu$ Eri $-$ $\xi$ Eri' prewhitened with 
0.615 d$^{-1}$, $2f'_{\rm orb}$ and $6f'_{\rm orb}$. Arrows indicate 
peaks at frequencies equal to $f'_{\rm orb}$, $3f'_{\rm orb}$, 
$4f'_{\rm orb}$, $5f'_{\rm orb}$ and $8f'_{\rm orb}$.}
 \label{Fig_2}
\end{figure}

In the first-run $u$ and $v$ power spectra, the highest peaks occurred 
at the same frequency of 0.615 d$^{-1}$. This frequency is close to 
$f'_1$, the only one found for $\mu$ Eri in Paper I (see the 
Introduction). The second and third runs yielded frequencies of 0.272 
and 0.815 d$^{-1}$, again the same for $u$ and $v$. The first of these 
numbers is close to twice the orbital frequency of $\mu$ Eri, $f'_{\rm 
orb}=0.13589$ d$^{-1}$, while the second, to six times this frequency. 
The low-frequency part of the power spectrum of the $v$ differential 
magnitudes `$\mu$ Eri $-$ $\xi$ Eri' prewhitened with 0.615 d$^{-1}$, 
$2f'_{\rm orb}$ and $6f'_{\rm orb}$ is shown in Fig.~\ref{Fig_2}. In 
this figure, the arrows indicate peaks at frequencies equal to $f'_{\rm 
orb}$, $3f'_{\rm orb}$, $4f'_{\rm orb}$, $5f'_{\rm orb}$ and $8f'_{\rm 
orb}$. Peaks at these frequencies are also present in the $u$ 
power-spectrum. 

\begin{figure}
\epsfxsize=240pt \epsfbox[52 52 107 117]{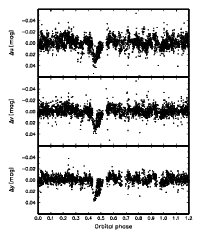}
 \caption{The $u$ (top), $v$ (middle) and $y$ (bottom) differential 
magnitudes `$\mu$ Eri $-$ $\xi$ Eri' prewhitened with $f'_{\rm 1}$ 
are shown as a function of orbital phase of $\mu$ Eri. Phase zero 
corresponds to HJD\,2452800.}
 \label{Fig_3}
\end{figure}

The occurrence of so many harmonics of the orbital frequency implies 
that the data contain a strongly non-sinusoidal signal of this 
frequency. The first possibility that comes to mind is an eclipse. 
Fig.~\ref{Fig_3}, in which the differential magnitudes `$\mu$ Eri $-$ 
$\xi$ Eri' prewhitened with 0.615 d$^{-1}$ are plotted as a function of 
orbital phase, shows that $\mu$ Eri is indeed an eclipsing variable.

Returning to frequency analysis of the comparison-stars data, we 
rejected observations falling in the orbital phase range from 0.4 to 
0.54, i.e., the data affected by the eclipse, and recomputed the 
power spectra. In the first-run $u$ and $v$ power-spectra, the highest 
peaks occurred at the same frequency of 0.615 d$^{-1}$ as before. The 
highest peaks in the second and third $u$ run were at 0.701 and 
0.813 d$^{-1}$, respectively, while the second and third $v$ runs 
yielded 1.206 and 0.701 d$^{-1}$. 

In the next step, we carried out nonlinear least-squares solutions 
separately for the $u$, $v$ and $y$ data. As starting values, we used 
all four frequencies found above, i.e., 0.615, 0.701, 0.8132 and 1.206 
d$^{-1}$. The results are presented in the first four lines of Table 4. 
The fifth frequency, $f'_5$, will be explained shortly. The frequencies 
and their standard deviations, listed in column 2, were computed as 
straight means from the separate solutions for the three bands. The 
amplitudes, $A_u$, $A_v$ and $A_y$, and their standard deviations, are 
given in columns 4, 5 and 6. The $v$-amplitude signal-to-noise ratio, 
computed in the same way as in Sect.\ 3.1, is listed in the last column. 

\begin{table*}
 \centering
 \begin{minipage}{130mm}
  \caption{Frequencies, periods and amplitudes in the out-of-eclipse 
differential magnitudes `$\mu$ Eri $-$ $\xi$ Eri' from the 2003-4 data. 
The last column contains the $v$-amplitude signal-to-noise ratio.}
  \begin{tabular}
{rl@{\hspace{3pt}$\pm$\hspace{3pt}}lr@{\hspace{3pt}$\pm$\hspace{3pt}}lr@
{\hspace{3pt}$\pm$\hspace{3pt}}
rr@{\hspace{3pt}$\pm$\hspace{3pt}}rr@{\hspace{3pt}$\pm$\hspace{3pt}}rr} \\
  \hline
  \hline
  ID &\multicolumn{2}{c}{Frequency [d$^{-1}$]} &\multicolumn{2}{c}
{Period [d]} & \multicolumn{2}{c}{$A_u$ [mmag]} & 
\multicolumn{2}{c}{$A_v$ [mmag]} & \multicolumn{2}{c}{$A_y$ [mmag]}&$S/N$\\
 \hline
 \hline
$f'_1 $&0.61504&0.00010&1.6259&0.00026&9.4&0.22&6.1&0.17&5.7&0.16&11.1\\
$f'_2 $&0.70160&0.00027&1.4253&0.00055&4.0&0.22&2.4&0.16&2.2&0.15&3.9\\
$f'_3 $&0.81351&0.00026&1.2292&0.00039&3.3&0.22&2.2&0.17&2.5&0.17&4.1\\
$f'_4 $&1.20739&0.00027&0.8282&0.00019&3.1&0.23&2.3&0.17&2.3&0.16&6.0\\
$f'_5 $&0.65934&0.00028&1.5167&0.00064&3.4&0.21&2.3&0.16&2.4&0.15&4.2\\ 
\hline
\hline
\end{tabular}
\end{minipage}
\end{table*}

\begin{figure}
\includegraphics{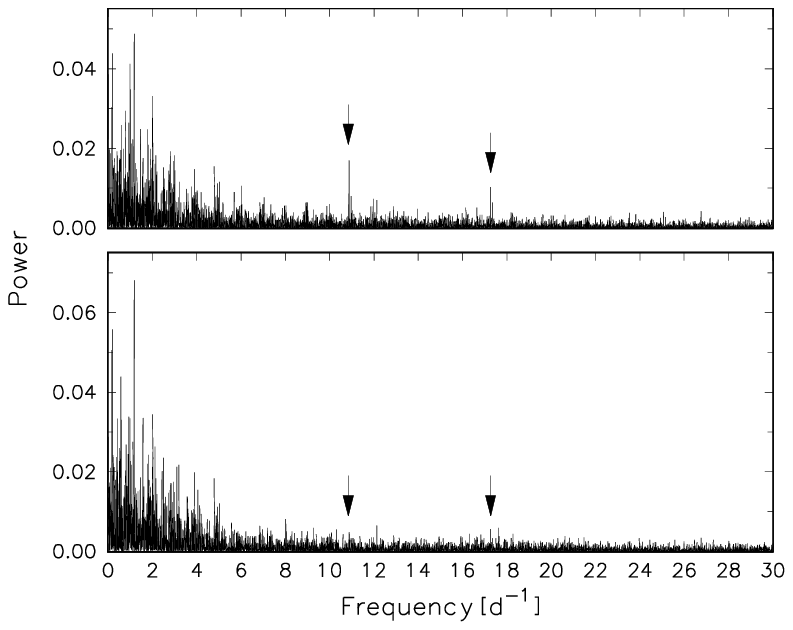}
 \caption{Power spectra of the $u$ (lower panel) and $v$ (upper) 
out-of-eclipse differential magnitudes `$\mu$ Eri $-$ $\xi$ Eri' 
prewhitened with the five frequencies of Table 4. Arrows indicate 
frequencies $f_{\rm x}=10.874$ and $f_{\rm y}=17.254$ d$^{-1}$.}
 \label{Fig_4}
\end{figure}

In the power spectra of the $u$ and $v$ differential magnitudes `$\mu$ 
Eri $-$ $\xi$ Eri' prewhitened with the first four frequencies of Table 
4, the highest peaks occur at the same frequency of 0.659 d$^{-1}$. 
Since a term of very nearly the same frequency is prominent in the 
frequency spectra of the 2002-3 comparison-stars data (see Sect.\ 4.2), 
we conclude that the signals at 0.659 d$^{-1}$ are intrinsic. A 
five-frequency nonlinear least-squares solution yielded the value of the 
fifth frequency and the corresponding amplitudes given in the last line 
of Table 4. 

The power spectra of the $u$ and $v$ differential magnitudes `$\mu$ Eri 
$-$ $\xi$ Eri' prewhitened with the five frequencies of Table 4 are 
shown in Fig.~\ref{Fig_4}. In both spectra the highest peak occurs at 
1.182 d$^{-1}$. Although these peaks may represent another term in the 
variation of $\mu$ Eri, we shall terminate the analysis at this stage 
for fear of over-interpreting the data.

At high frequencies, peaks at $f_{\rm x}=10.874$ and $f_{\rm y}=17.254$ 
d$^{-1}$ can be clearly seen in the $v$ spectrum (Fig.~\ref{Fig_4}, 
upper panel). The $v$-amplitude signal-to-noise ratio is equal to 5.2 
and 4.4 for $f_{\rm x}$ and $f_{\rm y}$, respectively. Although in the 
$u$ spectrum in the lower panel the peaks at these frequencies are 
masked by noise, a closer examination shows that they are present. For 
$f_{\rm x}$, the $u$, $v$ and $y$ amplitudes computed from the 
differential magnitudes `$\mu$ Eri $-$ $\xi$ Eri' amount to $0.9\pm0.22$, 
$1.2\pm0.16$ and $1.0\pm0.16$ mmag (formal sigmas), respectively. To 
within one formal $\sigma$, these numbers agree with the $f_{\rm x}$ 
amplitudes derived from the `$\nu$ Eri $-$ $\xi$ Eri' differential 
magnitudes (see Table 2). For $f_{\rm y}$, the $u$, $v$ and $y$ 
amplitudes computed from the differential magnitudes `$\mu$ Eri $-$ $\xi$ 
Eri' are equal to $0.9\pm0.22$, $0.9\pm0.16$ and $0.7\pm0.16$ mmag, 
respectively. Again, there is a $1\sigma$ agreement with the amplitudes 
obtained from the `$\nu$ Eri $-$ $\xi$ Eri' data (see Table 2). In 
addition, the phases agree to within $1\sigma$ in all cases. We conclude 
that both frequencies are due to an intrinsic variation of $\xi$ Eri.

\section{Low frequencies from the 2002-3 data}

\subsection{The eclipse}

\begin{figure} 
\epsfxsize=240pt \epsfbox[52 52 107 117]{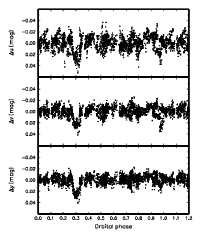}
\caption{The $u$ (top), $v$ (middle) and $y$ (bottom) 2002-3 differential 
magnitudes `$\mu$ Eri $-$ $\xi$ Eri' prewhitened with $f'_{\rm 1}$ are 
shown as a function of orbital phase of $\mu$ Eri. As in 
Fig.~\ref{Fig_3}, phase zero corresponds to HJD\,2452800.} 
\label{Fig_5} 
\end{figure}

We have to admit that in our original analysis of the 2002-3 data we 
missed the eclipse of $\mu$ Eri (see Paper I and the Introduction). 
Fig.~\ref{Fig_5} shows phase diagrams in which the 2002-3 differential 
magnitudes `$\mu$ Eri $-$ $\xi$ Eri' prewhitened with $f'_1$ are plotted 
as a function of orbital phase. The eclipse can be seen clearly.

A comparison of the phase diagrams in Figs.~\ref{Fig_5} and \ref{Fig_3} 
shows that while the middle of the eclipse in 2002-3 falls at a phase of 
about $0.30$, in 2003-4 it does at about $0.47$, indicating a problem 
with Hill's (1969) value of the orbital period. Assuming that Hill's 
value yields a correct cycle count between the first eclipse observed in 
2002 and the last one in 2003, we arrive at the following ephemeris: 
\begin{equation} 
{\rm Min.\ light}={\rm HJD}\,2452574.04\,(4) + E/0.135490\,(18), 
\end{equation} 
where $E$ is the number of cycles elapsed from the epoch given (which is 
that of the middle of the first eclipse we caught in 2002), and the 
numbers in parentheses are estimated standard deviations with the 
leading zeroes suppressed. The question why our photometric period differs 
from Hill's spectrographic one will be addressed in a forthcoming paper. 

\subsection{Analysis of the out-of-eclipse $\bmu$ Eri data}

In 2002-3 the numbers of differential magnitudes `$\mu$ Eri $-$ $\xi$ 
Eri' in the three bands were nearly the same, amounting to 2823, 2830 
and 2919 in $u$, $v$ and $y$, respectively. After we rejected 
observations falling within the eclipse, these numbers were reduced to 
2597, 2603 and 2688, still sufficient for analysis. Using these reduced 
data, we computed power spectra separately for $u$, $v$ and $y$. The 
first two runs yielded the same frequencies of 0.616 and 0.701 d$^{-1}$ 
in all three bands. These frequencies are very nearly equal to $f'_1$ 
and $f'_2$ of Table 4. In the third run, the highest peak in the $u$ 
power-spectrum was at 0.657 d$^{-1}$, while in the $v$ and $y$ 
power-spectra the highest peaks were at the same frequency of 1.207 
d$^{-1}$. The first of these numbers is close to $f'_5$, while the 
second is nearly identical with $f'_4$ (see Table 4). 

The fourth run was, however, a disappointment. In the $u$ power-spectrum, 
the highest peak occurred at 1.000 d$^{-1}$, while in the $v$ and $y$ 
ones, at 0.032 d$^{-1}$. Since neither of these frequencies is likely to be 
intrinsic, we did not attempt to compute fifth-run power spectra.

The four frequencies found above, i.e., 0.616, 0.701, 1.207 and 0.657 
d$^{-1}$ were used as starting values in a four-frequency nonlinear 
least-squares solutions. Justification for including $f'_4$ in the $u$ 
solution comes from the fact that a peak at this frequency is prominent 
in the fourth-run $u$ power-spectrum. Likewise, $f'_5$ was included in the 
$v$ and $y$ solutions because prominent peaks at this frequency can 
be seen in the fourth-run $v$ and $y$ power-spectra.

The results of the four-frequency solutions are given in Table 5 above 
the horizontal line. This table has the same format as Table 4. However, 
the $S/N$ (last column) was now computed from the $y$ data. 

The four-frequency solutions did not include $f'_3$. Since peaks at this 
frequency were present in the fourth-run $u$ and $v$ power-spectra, we 
carried out a five-frequency nonlinear least-squares solutions for all 
five frequencies of Table 4. The resulting values of $f'_3$, the 
amplitudes and $S/N$ are given in Table 5 below the horizontal line. 

\begin{table*}
 \centering
 \begin{minipage}{130mm}
  \caption{Frequencies, periods and amplitudes in the out-of-eclipse 
differential magnitudes `$\mu$ Eri $-$ $\xi$ Eri' from the 2002-3 data.
The last column contains the $y$-amplitude signal-to-noise ratio.}
  \begin{tabular}
{rl@{\hspace{3pt}$\pm$\hspace{3pt}}ll@{\hspace{3pt}$\pm$\hspace{3pt}}lr@
{\hspace{3pt}$\pm$\hspace{3pt}}
rr@{\hspace{3pt}$\pm$\hspace{3pt}}rr@{\hspace{3pt}$\pm$\hspace{3pt}}rr} \\
  \hline
  \hline
  ID &\multicolumn{2}{c}{Frequency [d$^{-1}$]} &\multicolumn{2}{c}
{Period [d]} & \multicolumn{2}{c}{$A_u$ [mmag]} & 
\multicolumn{2}{c}{$A_v$ [mmag]} & \multicolumn{2}{c}{$A_y$ [mmag]}&$S/N$\\
 \hline
 \hline
$f'_1 $&0.61587&0.00013&1.62372&0.00034&9.9&0.26&6.2&0.19&4.9&0.15&9.4\\
$f'_2 $&0.70143&0.00021&1.42566&0.00043&6.9&0.25&4.4&0.19&3.3&0.15&4.1\\
$f'_4 $&1.20690&0.00030&0.82857&0.00021&3.2&0.25&3.4&0.19&2.7&0.15&4.5\\
$f'_5 $&0.65751&0.00036&1.5209&0.0008&4.2&0.25&2.3&0.19&2.2&0.15&4.2\\
\hline
$f'_3 $&0.8147&0.0006&1.2275&0.0009&3.1&0.24&2.3&0.19&0.8&0.15&1.4\\
\hline
\hline
\end{tabular}
\end{minipage}
\end{table*}

\subsection{$\bnu$ Eri}

As explained in Paper I, the 2002-3 differential magnitudes of $\nu$ Eri 
were computed as `$\nu$ Eri $minus$ the mean of comparison 
stars,' but with the low-frequency variations of $\mu$ Eri filtered out. 
Thus, a peak at low frequency in the power spectrum of these 
differential magnitudes---if not caused by noise---would be due to an 
intrinsic variation of $\nu$ or $\xi$ Eri. In the latter case, however, 
the peaks would be suppressed by a factor of four, while the 
corresponding amplitudes would be divided by two. 

In Paper I we found a prominent peak at 0.254 d$^{-1}$ in the amplitude 
spectrum of the $v$ differential-magnitudes prewhitened with all 23 
frequencies identified from the 2002-3 data. The power spectrum of the 
same data also shows a prominent peak at this frequency. However, there 
is little power at this frequency in the 2003-4 spectra shown in 
Fig.~\ref{Fig_1}. On the other hand, in the 2002-3 $u$, $v$ and $y$ 
power-spectra there are peaks at 0.615 d$^{-1}$, a frequency very nearly 
equal to $f_{\rm B}$ found in Sect.\ 3.1 from the 2003-4 data. The 
2002-3 $u$, $v$ and $y$ amplitudes at this frequency amount to 
$2.5\pm0.19$, $1.2\pm0.14$ and $1.4\pm0.12$ mmag, respectively, in fair 
agreement with the 2003-4 amplitudes listed in Table 2. Since multiplying 
the 2002-3 amplitudes by two would make the agreement much worse, the 
possibility that $f_{\rm B}$ is due to $\xi$ Eri can be rejected. 

\section{Analysis of the combined data}

\subsection{$\bnu$ Eri}

\begin{table*}
 \centering
 \begin{minipage}{125mm}
  \caption{Frequencies and amplitudes in the differential magnitudes of 
$\nu$ Eri from the combined 2002-2004 data. Independent frequencies are 
listed in the upper part of the table. The combination frequencies are 
listed below the horizontal line. In both cases the frequencies are 
ordered according to decreasing $v$ amplitude, $A_v$. Frequencies 
$f_{\rm x}$ and $f_{\rm y}$ are due to $\xi$ Eri.}
  \begin{tabular}
{rl@{\hspace{3pt}$\pm$\hspace{3pt}}lr@{\hspace{3pt}$\pm$\hspace{3pt}}
rr@{\hspace{3pt}$\pm$\hspace{3pt}}rr@{\hspace{3pt}$\pm$\hspace{3pt}}rr} \\
  \hline
  \hline
  ID &\multicolumn{2}{c}{Frequency [d$^{-1}$]} & \multicolumn{2}{c}{$A_u$ [mmag]} & 
\multicolumn{2}{c}{$A_v$ [mmag]} & \multicolumn{2}{c}{$A_y$ [mmag]}&$S/N$\\
 \hline
 \hline
 $f_1   $      &\ \,5.7632828  &  0.0000019&72.8&0.13&40.8& 0.10& 36.7&0.10&214.7\\
 $f_2   $      &\ \,5.6538767  &  0.0000030&38.5&0.13&27.1& 0.10& 25.4&0.10&142.6\\
 $f_3   $      &\ \,5.6200186  &  0.0000031&35.0&0.13&24.5& 0.10& 23.2&0.10&129.0\\
 $f_4   $      &\ \,5.6372470  &  0.0000038&31.8&0.13&22.3& 0.10& 21.0&0.10&117.4\\
 $f_5   $      &\ \,7.898200   &  0.000032 & 3.6&0.13& 2.6& 0.10&  2.5&0.10&14.5\\
 $f_{\rm A}$   &\ \,0.432786   &  0.000032 & 4.1&0.13& 2.5& 0.10&  2.5&0.10&8.3\\
 $f_7   $      &\ \,6.262917   &  0.000044 & 2.8&0.14& 2.0& 0.10&  1.8&0.10&11.0\\
 $f_6   $      &\ \,6.243847   &  0.000042 & 3.0&0.13& 1.9& 0.10&  2.1&0.10&10.5\\
 $f_B   $      &\ \,0.61440   &  0.00005 & 3.0&0.13& 1.4& 0.10&  1.6&0.10&5.5\\
 $f_9   $      &\ \,7.91383   &  0.00008 & 1.7&0.13& 1.1& 0.10&  1.2&0.10&6.1\\
 $f_{\rm x}$    &  10.87424   &  0.00012  & 0.8&0.13& 1.0& 0.10&  0.7&0.10&5.7\\
 $f_{10}  $    &\ \,7.92992    & 0.00010  & 1.2&0.13& 0.9& 0.10&  0.9&0.10&5.0\\
 $f_8   $      &\ \,7.20090   &  0.00009 & 1.4&0.13& 0.9& 0.10&  0.9&0.10&5.0\\
 $f_{11}  $    &\ \,6.73223    & 0.00012  & 1.0&0.13& 0.8& 0.10&  0.6&0.10&4.5\\
 $f_{12}  $    &\ \,6.22360    & 0.00012  & 0.9&0.13& 0.8& 0.10&  0.8&0.10&4.4\\
 $f_{\rm y}$   &  17.25241 &  0.00016 & 0.6 & 0.13 & 0.6 & 0.10 & 0.5 & 0.10&4.4\\                             
\hline 
 $f_1+f_2    $  &  11.4171595  &  0.0000036&12.4&0.13& 8.8& 0.10&  8.4&0.10&50.9\\
 $f_1+f_3    $  &  11.3833014  &  0.0000036&10.8&0.14& 7.6& 0.10&  7.3&0.10&44.0\\
 $f_1+f_4    $  &  11.4005298  &  0.0000042&10.2&0.14& 7.2& 0.10&  6.8&0.10&41.7\\
 $2f_1      $  &  11.5265656  &   0.0000027& 4.4&0.13& 3.1& 0.10&  2.9&0.10&17.9\\
 $f_1+f_2+f_3$  &  17.037178  &   0.000005& 3.6&0.13& 2.5& 0.10&  2.3&0.10&18.1\\
 $f_2+f_3    $  &  11.2738953  &  0.0000043& 2.6&0.13& 1.5& 0.10&  1.3&0.10&8.7\\
 $f_1-f_2    $  &\ \,0.1094061  &  0.0000036& 2.6&0.13& 1.5& 0.10&  1.6&0.10&4.0\\
 $2f_1+f_2   $  &  17.1804423  &   0.0000040& 1.9&0.13& 1.5& 0.10&  1.3&0.10&10.9\\
 $2f_1+f_4   $  &  17.163813  &   0.000005& 1.7&0.13& 1.4& 0.10&  1.2&0.10&10.1\\
 $2f_1+f_3   $  &  17.1465842  &   0.0000041& 1.6&0.13& 1.3& 0.10&  1.1&0.10&9.4\\
 $2f_1+f_2+f_3$  &  22.800461  &   0.000005& 1.4&0.13& 1.0& 0.09&  0.8&0.10&8.3\\
 $2f_2      $  &  11.3077534  &   0.0000042& 0.8&0.14& 0.8& 0.10&  0.5&0.10&4.6\\
 $f_1+f_2-f_3 $  &\ \,5.797141   &   0.000005& 1.0&0.13& 0.8& 0.10&  0.8&0.10&4.2\\
 $f_1+f_5    $  &  13.661483  &   0.000032 & 1.1&0.13& 0.7& 0.09&  0.8&0.10&4.5\\
 $f_1+f_3+f_4 $  &  17.020548  &   0.000005& 0.9&0.14& 0.7& 0.10&  0.7&0.10&5.1\\
 $f_2+f_4    $  &  11.291124  &   0.000005& 0.8&0.14& 0.7& 0.10&  0.6&0.10&4.0\\
 $f_1+f_2+f_4 $  &  17.054406  &   0.000005& 0.8&0.14& 0.6& 0.10&  0.8&0.10&4.4\\
 $f_1-f_4    $  &\ \,0.1260358   & 0.0000042& 1.7&0.13& 0.6& 0.10&  0.8&0.10&1.6\\
 $f_1+2f_2   $  &  17.071036  &   0.000005& 0.6&0.14& 0.5& 0.10&  0.4&0.10&3.6\\
 $f_3+f_4    $  &  11.257266  &   0.000005& 0.5&0.14& 0.5& 0.10&  0.3&0.10&2.9\\
\hline
\hline
\end{tabular}
\end{minipage}

\end{table*}

After slight mean-light-level adjustments, the 2002-3 and 2003-4 
differential magnitudes of $\nu$ Eri were combined, separately for $u$, 
$v$ and $y$. The combined, 2002-4 data have the time base-line of 525.8 
d. The analysis of the 2002-4 differential magnitudes was carried out in 
the same way as that of the 2003-4 data (see Sect.\ 3). Sixteen 
independent and 20 combination frequencies could be identified from the 
power spectra. In all cases but two the yearly aliases were 
significantly lower than the central peak, so that there was no $\pm1$ 
y$^{-1}$ uncertainty. This was to be expected because the 2002-3 and 
2003-4 observing windows span as much as $0.43$ y each (see Sect.\ 2). 
The two exceptions were $f_6$ and $f_{12}$. They will be discussed later 
in this section. 

The 36 frequencies derived from the combined data are listed in the 
first column of Table 6. As in Tables 2, 4 and 5, the values of the 
independent frequencies and their standard deviations, given in column 
2, were computed as straight means from the separate solutions for $u$, 
$v$ and $y$. The combination frequencies, listed below the 
horizontal line, were computed from the independent frequencies 
according to ID in the first column; their standard deviations were 
computed from the standard deviations of the independent frequencies 
assuming rms propagation of errors. The amplitudes, $A_u$, $A_v$ and 
$A_y$, given together with their standard deviations in columns 3, 4 and 
5, respectively, are from the independent solutions for $u$, $v$ and 
$y$. The $v$-amplitude $S/N$, computed in the same way as in Sect.\ 3.1, 
is given in the last column. It can be seen that all independent 
frequencies meet the significance condition of Paper I. Among 
combination frequencies, this condition is not satisfied in two cases, 
viz., $f_3+f_4$ and $f_1-f_4$. 

In addition to frequencies derived from the 2003-4 data (see Table 2), 
Table 6 contains two further high frequencies due to $\nu$ Eri, viz., 
$f_{11}$ and $f_{12}$. The latter is close to that of one of several 
``possible further signals'' listed in Table 3 of Paper I and to  
frequency $\nu_7$ obtained in Paper III from radial velocities of the 
Si\,III triplet around 457 nm. Frequency $f_{11}$ is new. In order 
to make sure that this frequency is not due to $\xi$ Eri, we examined 
the 2002-4 out-of-eclipse differential magnitudes `$\mu$ Eri $-$ $\xi$ 
Eri' prewhitened with the six frequencies of Table 7 (see the next 
subsection). In the periodograms, there were no peaks at $f_{11}$; the 
highest peak in the vicinity, at 6.7168 d$^{-1}$, had the $v$ amplitude 
equal to about 0.4 mmag and $S/N<2.5$. Analogous tests with the 2002-3 
data also proved negative. 

\begin{figure} 
\includegraphics{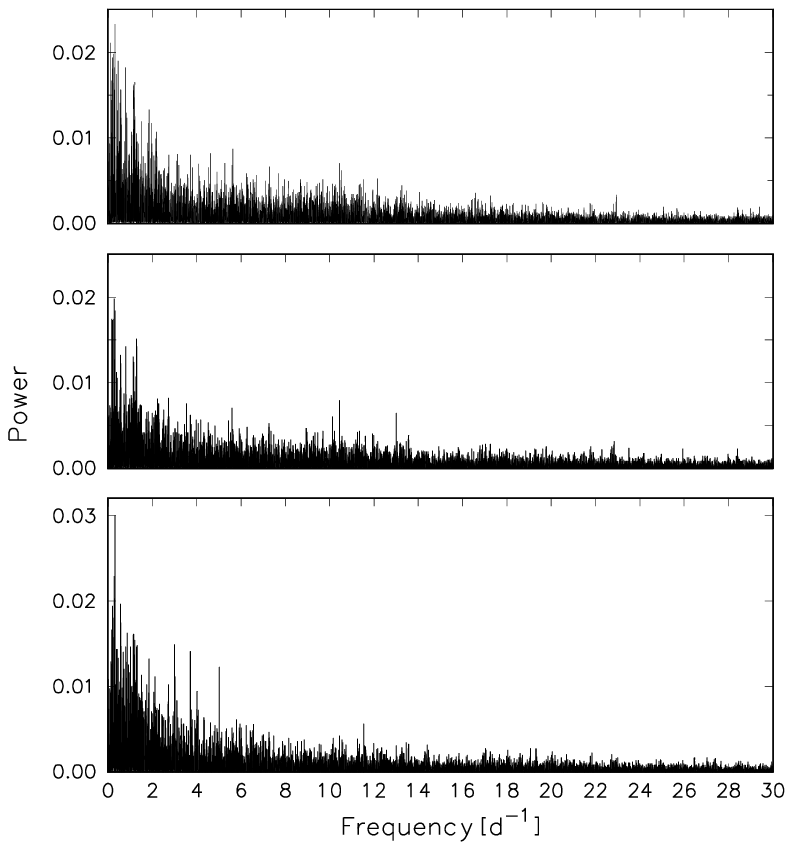} 
 \caption{Power spectra of the combined, 2002-4 $u$ (bottom),  $v$ 
(middle) and $y$ (top) differential magnitudes of $\nu$ Eri prewhitened 
with the 36 frequencies of Table 6.}
 \label{Fig_7}
\end{figure}

We shall now discuss the two problematic frequencies, $f_6$ and 
$f_{12}$, mentioned in the first paragraph of this section. In case of 
$f_6$, the central peak at $6.2438$ d$^{-1}$ was only slightly higher 
than the $+1$ y$^{-1}$ alias at $6.2465$ d$^{-1}$. However, the 2002-3 
and 2003-4 values of $f_6$ are both much closer to the frequency of the 
central peak than to that of the alias peak. Moreover, in each band, the 
nonlinear least-squares fit converged to exactly the same solution 
regardless of whether the starting value of $f_6$ was the frequency of 
the central peak, the 2002-3 value, or the 2003-4 value. We conclude 
that $f_6$ given in Table 6 is unlikely to be in error by $1$ y$^{-1}$. 

The case of $f_{12}$ is similar, but now the $-1$ y$^{-1}$ alias at 
6.2210 d$^{-1}$ is the problem. In $v$ and $y$ it is almost as high as 
the central peak at 6.2236 d$^{-1}$, while in $u$ it is even slightly 
higher. Computing power spectra from the averaged $u$, $v$ and $y$ 
residuals, with proper weight given to each band, did not solve the 
problem. In Paper III, there is also the y$^{-1}$ uncertainty: the 
frequency is equal to 6.22304 d$^{-1}$ for Si\,III 455.3 nm, but for 
Si\,III 456.8 and 457.5 nm it is close to 6.2210 d$^{-1}$. The 
frequency given in Table 3 of Paper I is equal to the alias frequency.  
More data are needed to decide whether the value given in Table 6 is the 
correct one. 

Fig.~\ref{Fig_7} shows the power spectra of the 2002-4 data prewhitened 
with the 36 frequencies of Table 6. In the $u$ and $y$ spectra (bottom 
and top panel, respectively) the highest peaks occur at 0.3142 d$^{-1}$, 
but in the $v$ spectrum, at 0.2625 d$^{-1}$. The corresponding 
amplitudes amount to 1.9 and 1.2 mmag in $u$ and $y$, and 1.1 mmag in 
$v$. In neither case does the signal-to-noise ratio exceed 3.4, so that 
these peaks are unlikely to be intrinsic. The reader may remember that 
in Sect.\ 3.1, a peak seen at 0.314 d$^{-1}$ in the 2003-4 $u$ power 
spectrum was also dismissed as spurious. 

The highest $S/N$ peaks in Fig.~\ref{Fig_7} have frequencies equal to 
5.0139 d$^{-1}$ in $u$ ($S/N=4.5)$, 13.0152 d$^{-1}$ in $v$ ($S/N=4.2)$ 
and 22.9440 d$^{-1}$ in $y$ ($S/N=4.1)$. The peak at 5.0139 d$^{-1}$ can 
be explained in terms of colour extinction in the $u$ band (see Sect.\ 
3.1). At 13.0152 d$^{-1}$, there are low peaks in the $u$ and $y$ 
spectra with $S/N$ equal to 2.9 and 2.4, respectively. No peaks at 
13.0152 d$^{-1}$ can be seen in the `$\mu$ Eri $-$ $\xi$ Eri' 
power-spectra, so that this frequency is not due to $\xi$ Eri. Finally, 
the frequency of 22.9440 d$^{-1}$ is very close to the combination 
frequency $3f_1+f_2$. We conclude that while the latter frequency may be 
intrinsic, the former is probably spurious.

\subsection{Out-of-eclipse variation of $\bmu$ Eri}

Combining the 2002-3 and 2003-4 out-of-eclipse differential magnitudes 
`$\mu$ Eri $-$ $\xi$ Eri,' we obtained time series consisting of 5818, 
5823 and 5189 data-points in $u$, $v$ and $y$, respectively. 
\begin{table*}
 \centering
 \begin{minipage}{140mm}
  \caption{Frequencies, periods and amplitudes in the out-of-eclipse 
differential magnitudes `$\mu$ Eri $-$ $\xi$ Eri' from the combined, 
2002-4 data. The last column contains the $v$-amplitude signal-to-noise 
ratio.}
  \begin{tabular}
{rl@{\hspace{3pt}$\pm$\hspace{3pt}}ll@{\hspace{3pt}$\pm$\hspace{3pt}}lr@
{\hspace{3pt}$\pm$\hspace{3pt}}
rr@{\hspace{3pt}$\pm$\hspace{3pt}}rr@{\hspace{3pt}$\pm$\hspace{3pt}}rr} \\
  \hline
  \hline
  ID &\multicolumn{2}{c}{Frequency [d$^{-1}$]} &\multicolumn{2}{c}
{Period [d]} & \multicolumn{2}{c}{$A_u$ [mmag]} & 
\multicolumn{2}{c}{$A_v$ [mmag]} & \multicolumn{2}{c}{$A_y$ [mmag]}&$S/N$\\
 \hline
 \hline
$f'_1 $&0.615739&0.000016&1.624065&0.000042&9.9&0.16&6.3&0.12&5.6&0.11&12.2\\
$f'_2 $&0.700842&0.000034&1.42686&0.00007  &4.8&0.16&3.0&0.12&2.5&0.10&6.5\\
$f'_4 $&1.205580&0.000043&0.829476&0.000030&3.0&0.16&2.6&0.12&2.4&0.10&7.2\\
$f'_{4{\rm a}}$*&1.208346&0.000042&0.827578&0.000029&2.9&0.16&2.5&0.12&2.4&0.10&6.9\\
$f'_5 $&0.658876&0.000039&1.51774&0.00009  &4.0&0.16&2.5&0.12&2.6&0.11&4.8\\
$f'_6 $&0.56797&0.00005&1.76066&0.00016  &2.7&0.16&2.0&0.12&2.1&0.11&3.1\\
$f'_3 $&0.81272&0.00006&1.23044&0.00009  &2.8&0.16&1.9&0.12&1.4&0.10&4.0\\
\hline
\hline
\end{tabular}
\flushleft
*)~$f'_{4{\rm a}}\approx f'_4+1\ {\rm y}^{-1}$\\
\end{minipage}
\end{table*}

The highest peaks in successive power-spectra of these data occurred at 
frequencies close to those found from the 2003-4 and 2002-3 time-series 
separately (see Tables 5 and 4) and at the frequency $f'_6=0.568$ 
d$^{-1}$, which is new. In $u$, the frequencies appeared in the order 
$f'_1$, $f'_2$, $f'_5$, $f'_3$ and $f'_{4{\rm a}}$, where the last 
frequency is the $+1$ y$^{-1}$ alias of $f'_4=1.2056$ d$^{-1}$. The two 
highest peaks in the sixth-run power spectrum occurred at 2.009 and 
0.997 d$^{-1}$, neither of which is likely to be intrinsic. The third 
peak, only slightly lower than the second one, was at $f'_6$. In $v$, 
the order was  $f'_1$, $f'_2$, $f'_4$, $f'_5$, $f'_6$ and $f'_3$, but in 
the third run the peak at $f'_4$ was only slightly higher than the one 
at $f'_{4{\rm a}}$. In $y$, the order was the same as in $v$, except 
that in the last run the highest peak occurred at 0.997 d$^{-1}$, while 
the peak at $f'_3$---although present---would be missed if it were not 
previously found in $u$ and $v$.

The $1$ y$^{-1}$ uncertainty which affects $f'_4$ did not plague other 
frequencies; in all other cases the yearly aliases were significantly 
lower than the central peak. 

The results of the analysis are given in Table 7. The numbers for 
$f'_{4{\rm a}}$ are from nonlinear least-squares fits in which the 
alias frequency read off the power spectrum was used as the starting 
value. In these fits, the other frequencies and the corresponding 
amplitudes were only slightly different from those given in the table. 

The $u$ to $y$ amplitude ratio in Table 7 exceeds 1.2 for all 
frequencies. For four frequencies the ratio is greater than 1.5, while 
for $f'_4$ and $f'_6$, it is equal to about 1.3. However, this dichotomy 
may be illusory because the (formal) standard deviation of the latter 
number amounts to about 0.10. 

The reader may have noticed that $f'_1 \approx f_{\rm B}$. Since neither 
frequency can be due to $\xi$ Eri because the amplitudes and phases do 
not match (see also Sect.\ 4.3), this curious near-equality must be an 
accidental coincidence. 

\section{Summary and clues for asteroseismology}

\subsection{Independent high frequencies of $\bnu$ Eri}

\begin{figure} 
\includegraphics{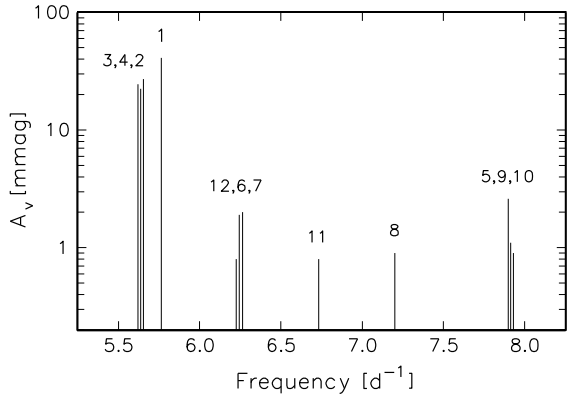} 
 \caption{Schematic $v$-amplitude spectrum of $\nu$ Eri from the 
combined, 2002-4 data: the 11 independent high-frequency terms, numbered 
as in Table 6.}
 \label{Fig_8}
\end{figure}

Fig.~\ref{Fig_8} shows schematically the 11 independent high-frequency 
terms of $\nu$ Eri derived from the combined, 2002-4 data. Comparing 
this figure with Fig.\ 4 of Paper I one can see that two of the three 
high-frequency ``possible further signals'' of Paper I are now upgraded 
to the status of certainty. This has already been mentioned in Sect.\ 
3.2 and 5.1. The terms in question are the $i=9$ and 12 ones. Both 
are members of close frequency triplets. 

The third high-frequency ``possible further signal'' of Paper I, with 
frequency equal to 7.252 d$^{-1}$, must remain in limbo. Although in 
the power spectra prewhitened with the 36 frequencies of Table 6 (see 
Fig.~\ref{Fig_7}) there is a series of low peaks in the vicinity of 7.25 
d$^{-1}$, the corresponding $v$-amplitudes are smaller than 0.6 mmag and 
the signal-to-noise ratios do not exceed 3.6. In $u$, the amplitudes are 
smaller than 0.8 mmag and $S/N<3.1$. More data are needed to decide 
whether any of these peaks is intrinsic. 

For the close frequency triplets seen in Fig.~\ref{Fig_8}, the mean 
separation, $S$, and the asymmetry, $A$, are listed in Table 8. 

\begin{table}
 \centering
 \begin{minipage}{77mm}
  \caption{The mean separation, $S$, and asymmetry, $A$, of the 
close frequency triplets in the oscillation spectrum of $\nu$ Eri.}
\begin{tabular}{cl@{$\pm$}ll@{$\pm$}l}
  \hline
Terms &\multicolumn{2}{c}{$S$ [d$^{-1}$]} &\multicolumn{2}{c}{$A$ [d$^{-1}$]}\\
\hline
3,4,2 &0.0169290&0.0000022&$-$0.000600&0.0000087\\
12,6,7 &0.019658&0.000064&$-$0.00118&0.00015\\
5,9,10&0.015860&0.00008&\hspace{7pt}0.00046&0.00019\\
  \hline
\end{tabular}
\end{minipage}
\end{table}

\begin{figure} 
\includegraphics{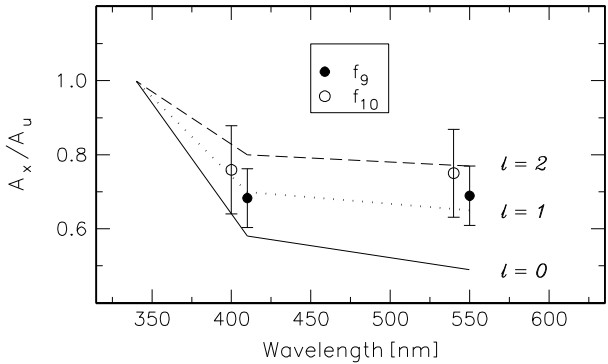} 
 \caption{Observed $uvy$ amplitude ratios for the small-amplitude members 
of the 7.91 d$^{-1}$ triplet (circles with error bars) compared with 
theoretical amplitude ratios for $\ell\leq2$. The observed ratios are 
from the 2002-4 amplitudes (see Table 6), the theoretical ones, from Fig.\ 5 
of Paper III. The open circles are shifted slightly in wavelength to avoid 
overlap.}
 \label{Fig_9}
\end{figure}

In Sect.\ 5.1 we have warned that $f_{12}$ may be in error by 1 
y$^{-1}$. If this were indeed the case, the values of $S$ and $A$ given 
in Table 8 for the 12,6,7 triplet should be replaced by $0.020964$ and 
$-0.00379$ d$^{-1}$, respectively. Because of this uncertainty, and the 
suspicion of a long-term variation of the amplitude of the $i=6$ term 
(Sect.\ 3.2), the triplet is not particularly suitable for 
asteroseismology at this stage. Fortunately, the other two triplets are 
well-behaved. There are no y$^{-1}$ uncertainties, even for the 
lowest-amplitude term of the 5,9,10 triplet, and no signs of long-term 
amplitude variation. In addition, the $\ell=1$ spherical harmonic 
identification for all members of the large-amplitude triplet and the 
$i=5$ member of the 5,9,10 one are secure (see Paper III or the 
Introduction). As can be seen from Fig.~\ref{Fig_9}, the $i=9$ and 10 
members of the triplet have $uvy$ amplitude ratios consistent with 
$\ell$ equal to 1 or 2. Unfortunately, the standard deviations of the 
amplitude ratios, especially those of the smallest-amplitude member, are 
too large to fix $\ell$ unambiguously. 

\subsection{Independent low frequencies of $\bnu$ Eri}

As we mentioned in the Introduction, the frequency resolution of the 
2002-3 data was insufficient to reject the possibility that $f_{\rm A}$ 
is equal to the combination frequency $3f_1-3f_3$. As can be seen from 
Table 6, the difference between $f_{\rm A}$ and $3f_1-3f_3$ amounts to 
0.0030 d$^{-1}$. This number is not only much larger than the 
standard deviation of $f_{\rm A}$, but also larger than the frequency 
resolution of the 2002-4 data by a factor of about 1.5. In fact, 
$3f_1-3f_3$ coincides with the $-1$ y$^{-1}$ alias of $f_{\rm A}$. Since  
the alias has about the same height as the $+1$ y$^{-1}$ one, the 
combination frequency's amplitude must be below the detection 
threshold. Consequently, there is no longer any doubt that $f_{\rm A}$ 
is an independent low frequency in the variation of $\nu$ Eri. In 
addition, we have found another low frequency, $f_{\rm B}$ (see Tables 2 
and 6). Since only high-order $g$ modes have frequencies that low, the 
suggestion put forward in Paper I that $\nu$ Eri were both a $\beta$ Cep 
variable and an SPB star is amply confirmed. 

\subsection{$\bmu$ Eri}

To the single frequency $f'_1=0.616$ d$^{-1}$, derived from the 
differential magnitudes `$\mu$ Eri $-$ $\xi$ Eri' in Paper I, we add 
five further ones (see Table 7). The values of the frequencies and the 
decrease of the $uvy$ amplitudes with increasing wavelength (for at 
least four frequencies) indicate that we are seeing high-radial-order 
$g$ modes. Thus, as already suggested in Paper I, the star is an SPB 
variable. Note that rotational modulation, the second hypothesis put 
forward in Paper I for explaining the $f'_1$ term is now untenable 
because it does not account for multiperiodicity. 

As can be seen from Table 7, the periods $P'_3$, $P'_2$ and $P'_1$ are 
equally spaced, with the spacing equal to $\sim\!0.20$ d, while $P'_4$ 
precedes $P'_3$ by twice this value. The equal spacing in period may be 
a manifestation of the well-known asymptotic property of high-order $g$ 
modes of the same $\ell$. There are, however, the following two problems 
with this idea: (1)~the term half-way between the $P'_3$ and $P'_4$ ones 
is missing, and (2)~the period-spacing is rather large. Better data may 
solve the first problem if the missing term is simply too weak to be 
detected in our data. The second problem requires a comparison with the 
theory. Unfortunately, the only SPB-star model available in the 
literature \citep*{dmp} has $M=4$ M$_{\sun}$, $\log L/{\rm 
L}_{\sun}=2.51$ and $X_{\rm c}=0.37$, whereas $\mu$ Eri is more massive 
(by $\sim\!2$ M$_{\sun}$), more luminous (by $\sim\!0.8$ dex), and more 
evolved (see Paper I). In the model, the largest period spacing (for 
$\ell=1$) is equal to $\sim\!0.07$ d, almost a factor 3 smaller than 
observed in $\mu$ Eri. Whether this disagreement can be alleviated with 
a model which better matches the star remains to be verified. If this 
turns out to be unsuccessful, one can still invoke the unlikely idea that 
an unknown amplitude limitation mechanism is suppressing the modes 
halfway between the observed ones. 

The possibility that instead of an equally-spaced period triplet,  
$P'_3$, $P'_2$, $P'_1$, we have a rotationally-split frequency triplet, 
$f'_1$, $f'_2$, $f'_3$, is much less likely. Indeed, for an $\ell=1$ 
$g$-mode with frequency equal to 0.7 d$^{-1}$ in the SPB-star model of 
\citet{dmp}, the observed mean separation of the frequency triplet, 
equal to $0.09849\pm0.00003$ d$^{-1}$, leads to equatorial velocity of 
rotation, $v_{\rm e}$, of about 30 km\,s$^{-1}$ (see their Fig.\ 13), 
whereas available estimates of $v_{\rm e} \sin i$ of $\mu$ Eri range 
from 150 to 190 km\,s$^{-1}$ (see Paper I). Increasing the model's 
radius in order to better match $\mu$ Eri may increase $v_{\rm e}$ to 
about 60 km\,s$^{-1}$, still much less than the observed values. 
An additional problem is posed by the large asymmetry of the frequency 
triplet. The asymmetry is equal to $0.0268\pm0.0001$ d$^{-1}$, while the 
rotational splitting seen in Fig.\ 13 of \citet{dmp} is nearly 
symmetric.  

In addition to confirming the SPB classification of Paper I, we have 
found $\mu$ Eri to be an eclipsing variable. As can be seen from 
Figs.~\ref{Fig_3} and \ref{Fig_5}, the eclipse is a transit, probably 
total, the secondary is fainter than the primary by several magnitudes, 
and the system is widely detached. As far as we are aware, the only 
other eclipsing variable with similar properties is 16 (EN) Lac 
\citep{j79}, except that in the latter case the eclipse is partial. 
(Interestingly, the discovery of an eclipse of this well-known $\beta$ 
Cephei variable was a by-product a three-site campaign undertaken for 
observing the star's pulsations.) Solving the $\mu$ Eri system will 
yield the primary's mean density and its surface gravity. This, however, 
is beyond the scope of the present paper. 

\subsection{$\bxi$ Eri}

The frequencies $f_{\rm x}=10.8742$ and $f_{\rm y}=17.2524$ d$^{-1}$ 
(see Tables 2 and 6) and the MK type of A2\,V (see the Introduction) 
leave no doubt that the star is a $\delta$ Scuti variable. The 
Str\"omgren indices, $c_1=1.076$ and $b-y=0.038$ \citep{hm}, are not 
reddened. This is not inconsistent with the star's Hipparcos parallax of 
$15.66\pm0.80$ mas. Using the parallax and the $V$ magnitude from 
\citet{hm} one gets $M_{\rm V}=1.12 \pm 0.11$, a  value which, together 
with the $b-y$ index, places the star about 0.02 mag to the blue of the 
observational blue edge of the $\delta$ Scuti instability strip in the 
$M_{\rm V}$ vs.\ $b-y$ diagram (see, e.g., \citealt{h02}). Apart from 
indicating the need for a slight revision of the blue edge, this 
position in the diagram suggests marginal pulsation driving as a 
possible explanation for the small $uvy$ amplitudes. However, in view of 
the star's high $v\sin i$ of 165 km\,s$^{-1}$ \citep{am}, another 
explanation may be provided by the hypothesis of \citet{b82} that fast 
rotation is a factor in limiting pulsation amplitudes. 

Unfortunately, with only two small-amplitude modes the asteroseismic 
potential of $\xi$ Eri is insignificant. 

\section*{Acknowledgments}

MJ and AP's participation in the campaign was supported by KBN grant 
5P03D01420. MJ would also like to acknowledge a generous allotment of 
telescope time and the hospitality of Lowell Observatory. GH's work was 
supported by the Austrian Fonds zur F\"orderung der wissenschaftlichen 
Forschung under grant R12-N02. ER thanks for the support by the Junta de 
Andalucia and the Direccion General de Investigacion (DGI) under project 
AYA2003-04651. The referee, Dr C. Simon Jeffery, helped us to improve 
the paper.

\bsp

\label{lastpage}

\end{document}